\documentclass{aa}
\usepackage{psfig}
\def\mnras{MNRAS} 
\def\aap{A\&A}
\def\apj{ApJ}
\def\apjl{ApJL}
\def\aj{AJ}
\def\apjs{ApJS}
\def\nat{Nat}
\begin{document}

\thesaurus{12(12.03.3,12.05.1,11.17.4APM08279+5255, 02.14.1)}

\title{ Low deuterium abundance in
the $z_{\rm abs}$=3.514 absorber towards  APM 08279+5255
\thanks{The data presented herein were obtained
at the W.M. Keck Observatory, which is operated as a
scientific partnership among the California Institute
of Technology, the University of California and
the National Aeronautics and Space Administration. The Observatory
was made possible by the generous financial support of the W.M. Keck
Foundation.}
}
\titlerunning{Deuterium towards  APM 08279+5255}
\authorrunning{Molaro et al}

\author{Paolo Molaro\inst{1} \and 
Piercarlo Bonifacio\inst{1} \and 
Miriam Centurion\inst{1,2} \and Giovanni Vladilo\inst{1}}

\offprints{P. Molaro}

\mail{molaro@oat.ts.astro.it}

\institute{ Osservatorio Astronomico di Trieste, Via G.B. Tiepolo 
11, 34131, Trieste \and  
Instituto de Astrof\'\i sica de Canarias, 38200 La Laguna,Tenerife, Spain}

\date{received .../Accepted...}

\maketitle

\begin{abstract}
{A high-resolution,  high signal-to-noise HIRES-Keck spectrum 
of  APM 08279+5255 
reveals a  
feature in the  HI Ly$\alpha$  profile 
of  the $z_{\rm abs}$=3.514 absorber 
at the  expected position of the  corresponding DI Ly$\alpha$ line. 
The absorber shows a relatively simple velocity structure, with a 
major  component detected in several metallic lines including
CII, CIV, SiII, SiIII and Si IV.
Modeling of the hydrogen column density
with the minimum number of components  yields  
$\frac{D}{H}\approx 1.5 
\times 10^{-5}$. 
However, a more complex structure for the 
hydrogen cloud with 
somewhat  ad hoc components would allow    a higher  $\frac{D}{H}$ .
The system has a very low metallicity with 
[Si/H] $\approx -3.5$ 
and [C/H] $\approx -4.0 $
and is therefore representative of  essentially 
unprocessed 
material, with a deuterium abundance close to  the primordial value.
The 
present analysis  
favours the low  
D/H value of $\frac{D}{H}$ =$3.39\pm 0.25 \times 10^{-5}$
measured towards 
QSO 1009+2956 and QSO 1937-1009   
\nocite{burles} (Burles \& Tytler 
1998a, 1998b )\nocite{burles2} as the fiducial
value of deuterium at high redshift.
}
\end{abstract}


\section{Introduction}

Deuterium,  together with  $\rm ^{3,4}He$ and $^7$Li 
is one of the few elements produced by nuclear reactions
in the first minutes after big bang \nocite{wfh}
(Wagoner, Fowler \& Hoyle 1967), with the difference that for deuterium 
SBBN is the only significant  source. 
Deuterium yields are the most sensitive to
the nuclear density at the nucleosynthesis epoch  making  deuterium the
most sensitive baryometer among the primordial light elements.
 
Deuterium 
is measured in the LISM and  extrapolation to  the primordial value 
requires    a  modeling
of the Galactic chemical evolution. The intrinsic uncertainties
in this extrapolation   can
be avoided by direct deuterium observations of almost primordial material
in the 
high redshift/unevolved  absorption systems  
 (Adams, 1976)\nocite{A76}. 
This was in fact achieved only  recently for a few systems,
but with
conflicting results and abundances which  differ by almost
one order of magnitude.
At present the number of systems analyzed in detail
is evenly distributed between two possibilities. Two systems provide
a {\it high}  D/H value at $\approx$ 2 $\times 10^{-4}$ 
(Songaila et al 1994\nocite{S94}, 
Carswell et al 1994\nocite{C94}, Webb et al 1997\nocite{Webb})
and other two systems a  {\it low} abundance at 3.39 ($\pm 0.25)\times 10^{-5}$ 
(Burles and Tytler
1998a,1998b). 

It has been pointed out that  the high deuterium case 
is always prone to  the  possibility
of a contamination of the 
deuterium line with an 
Ly$\alpha$ interloper. 
For the lower deuterium case 
either the hydrogen column density may have been overestimated or a 
special chemical evolution,  with large astration of D, may
have   occurred in these systems. 
The handful of detections found so far 
does not 
allow a choice  between  the different
cases  on  the basis of simple statistics.
The paucity of suitable systems is 
due to the fact that only a narrow range of HI column densities
allows  the detection of the DI line. At too low densities the DI line
is too weak for detection, whereas at too high densities the line is washed
out by the saturation of the HI line. Moreover, a suitable system must have a
rather simple velocity structure, ideally  a single cloud system, which is
obviously rare.  
Burles \& Tytler estimated that only about 1 out of 30 QSO's
satisfies these conditions.
Here we present a new system which shows these  characteristics  
 and  a measure of deuterium which 
supports the case for
low primordial deuterium.

\begin{figure}
\psfig{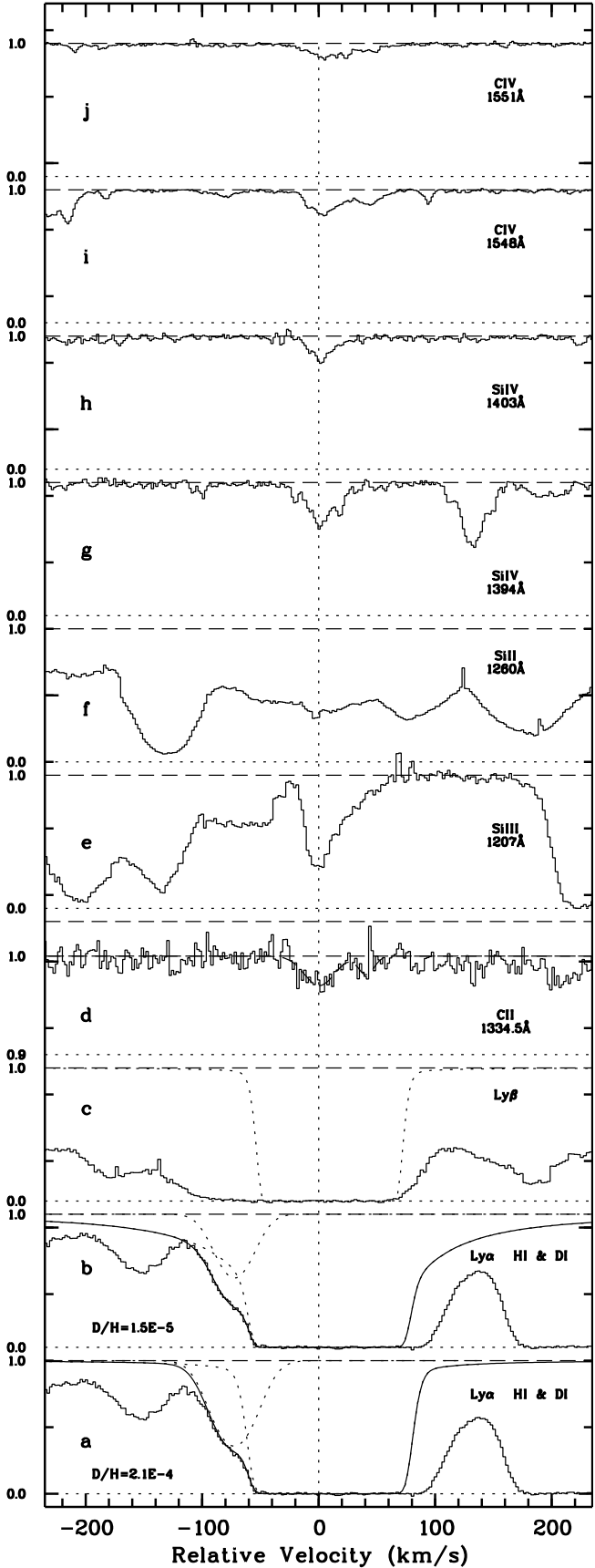}
\caption{Absorption system at z$_{\rm abs}$= 3.51374. See text for details. }
\label{fig1}
\end{figure}

\section{Observations and analysis}

The spectra  of APM 08279+5255 have been obtained
with the Keck I telescope and the HIRES spectrograph by 
\nocite{E99}Ellison et al (1999)
and made available for the astronomical community.
\footnote{\tt ftp://fttp.ast.cam.ac.uk/pub/papers/APM08279} An 
analysis of these spectra with details on the reduction procedure
and a log of observations 
can be found in the original paper. 
This QSO is one of the brightest
known also by virtue of gravitational lensing  magnification. 
The image of
this QSO reveals  two components of similar
intensity 
separated by about $0''.4$    
(Irwin et al 1998).\nocite{Irwin}

Ellison et al (1999)\nocite{E99} 
describe  the absorber at 
$z_{\rm abs}$=3.514 and identify   
the high ionization species of
SiIV and CIV associated with  the system.
A total of five  components are resolved  in SiIV and two in CIV, with
both ions showing  a main   component at redshift z$_{\rm abs}=3.5138$. 

Our attention was caught by the feature on the blue wing
of the Ly$\alpha$ profile of the 
displayed in Fig. 4 of Ellison et al 
\nocite{E99}, which  we suspected to be a D line.
In addition   we
have identified features 
due to CII 1334.5 \AA, SiII 1260.4 \AA\ and  SiIII 1206.5 \AA
(see Fig. 1).    
The CII and SiII lines are  very faint and the detection has been possible 
thanks to the quality
of the HIRES spectrum at this wavelength (S/N $\sim$ 125).
The SiII 1193.3 \AA\ absorption may also be  present, 
albeit strongly blended with a lower redshift
Ly$\alpha$ absorber. 
Ly$\beta$ falls in the available spectral range but is, unfortunately,
severely blended with local Ly$\alpha$ clouds  
(Fig. 1, panel $c$).

The CII profile shows a double component with the main absorption
occurring at redshift z$_{\rm abs}$=3.51374, namely the redshift at which the main 
features of CIV (z$_{\rm abs}=3.51380$) and SiIV (z$_{\rm abs}=3.51376$) also
occur, as measured by  
\nocite{E99}Ellison et al (1999). 
A third CII 1334 \AA\  absorption at about +200 km s$^{-1}$ is associated
to another Ly$\alpha$ cloud redwards of our system.
The CII and SiII ions are  better tracers of  neutral hydrogen (and deuterium)
than higher ionization species in Lyman limit systems
(\nocite{proch}Prochaska, 1999).

\begin{table}
\caption{Column densities.} 
\begin{tabular}{lcclc}
\hline
Element &$\lambda_{\rm lab}$(\AA) &{$z_{\rm abs}$}&
{log N(X)}  &{ b} \\
\hline
CII & 1334.5323 & 3.51374 & $12.24 \pm 0.05$ & $17.8 \pm 1.8$ \\
CII & 1334.5323 & 3.51435 & $11.85 $& $\rm ^{L}$ \\
\\
SiII & 1260.4221 & 3.51368 & $11.67$ & $\rm ^{L}$ \\
\\
SiIII & 1203.4221 & 3.51376 & $12.77$ & 14.7$\pm 0.5$ \\
\\
DI  & 1215.3394 & 3.51387 & $13.27 \pm 0.08$ & $21.2 \pm 2.3$\\
HI  & 1215.6700 & 3.51387 & $18.09 \pm 0.03$ & $21.0 \pm 0.6$\\
\hline
\end{tabular}
{$\rm ^L$ Linear part of the curve of growth\hfill}
\end{table}

To fit the observed profiles we used the {\tt fitlyman } package 
(Fontana \& Ballester 1995)
 \nocite{fontana}
in {\tt MIDAS}. The results for the analysis  are 
reported in Table 1 and the  fits are   shown in  
Fig. 1.
We  model the Ly$\alpha$ absorption profile assuming a dominant component in HI
as seen in all the metallic lines. Since the D line affects the 
blue wing of the absorption profile we 
restricted the fit to  a 
narrow window
($-125.5 , +28.1$
km\,s$^{-1}$) 
which includes only the blue wing. In the fitting  
procedure the 
central wavelength, 
broadening and column density of the single component
with the associate D are left as free parameters 
with the only restriction that z$_{\rm abs}$  must be the same for DI and HI. 
The best fit ($\chi^{2}$=0.28)
gives a 
deuterium abundance of $\frac{D}{H}$=$1.5\pm 0.3_{\rm stat}\times10^{-5}$ .
The statistical error follows from the fit and reflects only the data
quality; as discussed below, the error budget 
is  dominated by systematics related to the cloud modeling.
The column densities,  $b$ parameters,  z$_{\rm abs}$  and  1$\sigma$ errors
are given in Table 1.
The resulting redshift  
(z$_{\rm abs}$ = 3.51387) is
in agreement with the most intense
component of the metallic ions, 
thus supporting the identification of deuterium.
Fig. 1, panel $b$, shows the data and the fitted profile (thin solid line).
The dotted lines show the individual synthetic profiles of DI and HI.

Remarkably, the HI and DI  
broadening velocities are 
consistent with 
the $b$ value
of the CII 1334 \AA ~ absorption of the same system. 
This suggests that the HI, DI and CII absorptions are actually tracing
the same gas, where  
macroscopic, rather than thermal,  motions dominate the
line broadening. 
The broadening for the metal lines is larger than usually seen
in blend-free absorbers towards QSOs.
An additional source of
broadening may be  the lensing, since  
both images 
of the lens, separated by about $0''.4$ are
present in the spectrograph slit ($0''.86$)and recorded simultaneously. 
\par
The $\frac{D}{H}$ 
abundance derived here should be  
considered as a lower limit since we know  
from Si IV and C IV  that other  components 
might be present.
Adding new  components to the fit, although of lower column densities,
would inevitably slightly lower the H I column density for the major
component. 
Additional  components,  at slightly higher redshift, 
are required
to reproduce  the extra 
absorption on the red wing of Ly$\alpha$.
In any case this red 
component is  much smaller than 
the dominant one since its Ly$\beta$ is much weaker.
 
To illustrate how these effects can be important in 
Fig. 1, panel $a$, we show  a model for the Ly$\alpha$ line
with a lower hydrogen column density (
log N(HI) = 17.20) and a slightly higher deuterium
column density (log N(D I) = 13.52),
which corresponds to a ratio of $\frac{D}{H}$=2.1$\times 10 ^{-4}$.
This fit has been performed in a smaller window 
($-105.3, +28.1$ km\,s$^{-1}$) and  the 
$\chi^{2}$ is 0.35, i.e. comparable to the former case. 
However the fit
fails to reproduce the edge of the blue wing shortwards 
the deuterium feature, because of a less prominent
hydrogen Ly$\alpha$ damping wing. An additional
hydrogen cloud with relatively low column density
present at this
position  is required to provide a good fit to the entire absorption.
Since this solution requires an ad hoc absorption at the 
right place we
consider the former solution  more probable. 
Unfortunately Ly$\beta$ cannot help to further
constrain the H I column density. 
In  Fig. 1, panel $c$,
the observed Ly$\beta$ 
is shown together with the synthetic spectrum computed with
the parameters derived from  Ly$\alpha$ fitting to show that consistency is
achieved with  the predicted 
Ly$\beta$  sitting in a strongly
absorbed region. The  blue wing  of Ly$\beta$
is  clearly contaminated by  lower redshift Ly$\alpha$ interlopers 
and carries  no useful  information on the 
absorption system.
Observations of higher order terms of the Lyman series, if free from
blends, will certainly be important 
to further constrain the H I column density.

\section{Discussion}

Deuterium measurements in high redshift  absorbers of low
metallicity provide
the best estimate of the primordial deuterium. 
Relatively high deuterium ($
\frac {D}{H} \approx 2  \times  10^{-4}$)
has been measured  in  the system  at redshift z$_{\rm abs}$=3.32 
towards  QSO Q0014-813 (Songaila et al 1994\nocite{S94}, Carswell et al
1994)\nocite{C94}.
However the discovery of  a Ly$\alpha$ contaminant led 
Burles Kirkman and Tytler (1999)\nocite{bkt99}
to argue against the possibility to measure  D/H in this system.
A similarly high D/H value has been found by
Webb et al (1997)\nocite{Webb} in the system at z$_{\rm abs}$=0.7 
towards QSO 1718+4807 
(but see Tytler et al 1999).\nocite{t99} 
Other tentative and somewhat  more uncertain  detections, 
all giving high $\frac{D}{H}$  of $\approx 10^{-4}$,  have been reported 
by Carswell et al (1996)\nocite{C96} and Wampler et al 
(1996)\nocite{W96}. 
In contrast with  the high values of $\frac{D}{H}$, 
Burles \& Tytler(1998a,1998b)     
found $\frac{D}{H}$ about one order of
magnitude lower  in two Lyman Limit Systems   
towards Q1937-1009  and   Q1009+2956,
respectively. 
Both systems
have   similar  abundance, with an average of
$\frac{D}{H}$ =$3.39(\pm 0.25) 
\times 10^{-5}$.  
Levshakov et al (1998)\nocite{lev98}     derived
$\frac{D}{H}\approx 4.1 - 4.6 
\times 10^{-5}$   
by using the same observational data 
of Tytler and collaborators by accounting for spatial 
correlations in the large scale velocity field.
A new upper limit of  D/H $<6.7\times 10^{-5}$
in the Lyman limit system at $z=2.799$
towards QSO 0130-4021 has been
recently presented by Kirkman et al (1999).\nocite{kirk}

In this letter  we have shown  that the 
feature on the blue wing of the HI Ly$\alpha$ absorption
at $z_{\rm abs}=3.514$ in the spectrum of APM 08279+5255 can  be
identified  with confidence as the 
DI line of the main component of the absorber.  
The absorber is  an excellent
candidate  for the deuterium analysis
since it shows a  relatively simple structure, the appropriate
column density and a low metallicity. 
The [CII/HI] 
abundance 
\footnote{Using the standard 
definition [X/H]= log (X/H) - log (X/H)\sun}
derived for the system is  $-$2.4,
and the 
[SiII/HI] is $-1.97$ using the solar  abundances 
of Anders and Grevesse (1989)
\nocite{ag89}. 
However ionization corrections are not negligible.
 From the column densities of 
Si II, Si III and Si IV, a CLOUDY 
\nocite{fer}(Ferland, 1995) 
photoionization model computed by assuming    N(HI)
$ = 10^{18.1}\rm cm^{-2}$ and a Haardt-Madau
spectrum  at redshift 3.5
\nocite{hm}(Haardt \& Madau, 1996)  provides [Si/H]$=-3.5$ 
and log U $=-2.0$ 
(Burles, private communication).  
Carbon provides a worse fit, but with the optimal solution
obtained for  [C/H]$=-4.0$, consistent with the
typical overenhancement with respect to Si
observed in the Pop II Galactic halo stars.

A matter of concern in the analysis of the spectrum is the possible
influence of the lensing on the derived abundances
(see for instance Rauch et al 1999\nocite{rauch}). 
The possibility than only one light path intersects the absorber
is ruled out by the fact that the core of the Ly$\alpha$ absorption 
is at  zero intensity level (Fig. 1).
The transverse size  at $z=3.514$ depends on the redshift
of the lens, which is not known. Candidates are 
the damped system at $z=3.07$,
and the Mg II systems at $z=1.81$ and $z=1.18$ (Irwin et al, 
1998).\nocite{Irwin}
If the lens is the damped system
the transverse size is $\simeq 0.99$ kpc~$h_{50}^{-1}$
($\Omega_0 = 1$). This size  becomes  
as small as $\simeq 0.11$ kpc~$h_{50}^{-1}$
if the lens is at $z=1.18$.
The widths of metal lines set an upper limit  
only $\approx 10$ km\,s$^{-1}$
on the velocity
difference between the slabs  sampled by the two
light paths.
Therefore a small transverse size,
with the lens at $z=1.18$, is favoured, 
in agreement with the suggestion of Irwin et al (1998).
\nocite{Irwin} 
Since intrinsically large variations in D/H between different regions
are unlikely and we have no evidence of large
variations in radial velocities, the
D/H abundance analysis should not be affected
by the lensing.

Our derived $\frac{D}{H}$  abundance is of 
1.5 $ \times 10^{-5}$, about a factor of two
lower than the value of Burles \& Tytler (1998a, 1998b).
The case discussed here  supports  the low deuterium value
as the fiducial abundance in high redshift systems. 
The present value is 
of lower accuracy of the cases discussed by  
Burles and Tytler (1998a,1998b)
considering possible large systematic errors
in the hydrogen column density.  Larger values
are not completely ruled out and 
the measure may be considered  as a 
lower limit.
Considerations based on the consistency among the various
primordial elements also suggest a deuterium value somewhat higher
of what  is  measured here.
From BBN calculations 
performed
with the Kawano (1992\nocite{kawano}) code by 
assuming 3 massless neutrino families,
the baryon to photon ratio  ($\eta = n_b/n_\gamma$) implied by   
$\frac{D}{H}$ $= 1.5\times 10 ^{-5}$
is of  $8.4\times 10^{-10}$ .
This $\eta$  implies 
a helium abundance (in mass) of  Y = 0.251  and a lithium abundance of
$\rm \frac{^7 Li}{H}= 
8.3\times 10^{-10}$.
Making our D measurement 
consistent  with the Li
plateau value of $\simeq 1.7\times 10^{-10}$
requires    a factor of $\simeq 5$ Li depletion 
in Pop II dwarf stars.  This depletion
is greater than presently allowed
either by observations (Bonifacio \& Molaro 1997)\nocite{bm}
or theory (Michaud \& Charbonneau 1991)\nocite{mich}.
The high value implied for  Y  is also not
compatible with either the determinations of Pagel et al
\nocite{pagel}  (1992)
or  Izotov \& Thuan\nocite{izo98} (1998) and there have been no other claims
of  a primordial He in excess of these   values.

Most of the uncertainty in the present  measure  of $\frac{D}{H}$ 
is due to the uncertainty 
in the hydrogen column density,
which  is constrained only by the Ly$\alpha$ line. 
Additional observations
of the QSO at shorter wavelengths are needed to detect higher
members of the Lyman series. These may  allow to further constrain
the H I column density making possible in the future 
a more valuable deuterium measurement in the $z_{\rm abs}=3.514$ absorber.

\acknowledgements{ We are grateful to Ellison et al
for making their excellent data available to the community.
We thank the referee S. Burles for many valuable comments which
improved the paper.
We would also like to thank Elisabetta Caffau for 
performing the BBN calculations and Massimo Ramella
for helpful discussions on gravitational lenses. }

\end{document}